\documentclass[rmp,aps,preprint,nofootinbib,endfloats]{revtex4}

\usepackage{graphics}
\usepackage{epsfig}

\def\inbar{\,\vrule height1.5ex width.4pt depth0pt}
\def\IR{\relax{\rm I\kern-.18em R}}
\def\IC{\relax\hbox{$\inbar\kern-.3em{\rm C}$}}

\begin{document}

\title{Measuring and Understanding the Universe}

\author{Wendy L. Freedman}
\email{wendy@ociw.edu}
\affiliation{Observatories of the Carnegie Institution of Washington,
813 Santa Barbara St., Pasadena, CA 91101, USA}
\author{Michael S. Turner}
\affiliation{Center for Cosmological Physics and the Departments
of Astronomy \& Astrophysics and of Physics, The University of Chicago,
5640 S. Ellis Ave., Chicago, IL  60637-1433 and NASA/Fermilab Astrophysics
Center, Fermi National Accelerator Laboratory, MS 209, PO Box 500,
Batavia, IL  60510-0500}

\begin{abstract}
Revolutionary  advances in  both theory  and technology  have launched
cosmology   into  its   most   exciting  period   of  discovery   yet.
Unanticipated  components of  the universe  have been  identified,
promising ideas  for understanding the basic features  of the universe
are being tested, and deep connections between physics on the smallest
scales and on the largest scales are being revealed.
\end{abstract}

\maketitle
\tableofcontents

\section{Introduction}

Thirty years ago, cosmology was described as a search for two numbers:
the current expansion rate (or Hubble constant), $H_0$, and its change
over time, the deceleration parameter, $q_0$ (\onlinecite{Sandage:1970}).  But that
was  before the  discovery of  giant  walls of  galaxies, voids,  dark
matter, tiny variations in the  cosmic microwave background  radiation (CMB), dark
energy and the  acceleration of the Universe.  Today,  the subject has
become vastly richer,  and the numbers being sought  are more numerous
but more closely  tied to fundamental theory.  An  overall picture has
emerged that accounts for the  origin of structure and geometry of the
Universe, as  well as  describing its evolution  from a fraction  of a
second onward.

In this  new and still-evolving picture rooted  in elementary particle
physics, in  a tiny fraction of  a second during the  early history of
the universe, there was  an enormous expansion called inflation.  This
expansion  smoothed  out  wrinkles  and  curvature in  the  fabric  of
spacetime, and  stretched quantum fluctuations on  subatomic scales to
astrophysical  scales.   Following  inflation  was a  phase  when  the
Universe was  a hot  thermal mixture of  elementary particles,  out of
which arose  all the  forms of matter  that exist today.   Some 10,000
years into its evolution, gravity  began to grow the tiny lumpiness in
the  matter distribution  arising from  quantum fluctuations  into the
rich  cosmic structures seen  today, from  individual galaxies  to the
great clusters of galaxies and superclusters.

Recent observations of the universe have not only strengthened and expanded
the big-bang model, but they have also revealed surprises.  In
particular, most  of the universe  is made of  something fundamentally
different from the ordinary matter we  are made of.  (By ordinary
matter, we mean matter made of neutrons and protons; the jargon for
this is baryons, the technical term for particles made of quark
triplets.)  About 30\% of the
total mass--energy  is dark matter, composed of  particles most likely
formed early  in the universe.  Two thirds  is in a
smooth ``dark  energy'' whose gravitational effects
began causing the expansion of the universe to speed up just
a few billion  years ago.  Ordinary matter, the bulk  of it dark, only
accounts for  the remaining 4\% of  the total mass-energy  density of  the universe.
While the remnant (thermal) microwave background from the hot big bang
contributes  only  about  0.01\%,  it encodes  information  about  the
spacetime  structure  of the  Universe,  about its  early history,
and possibly even about its ultimate fate.

We  have also learned  much about  the organization  of the
universe.   In the  nearby  universe, galaxies  are  distributed in  a
``cosmic web''  composed of sheets and  sinuous filaments interspersed
with  voids (see  Figures 1 and 2).  Though  inhomogeneous on
these apparently
vast  scales, the  Universe  becomes more  and  more homogeneous  when
viewed on even larger scales  from 100\,Mpc out to the current horizon
of 10,000\,Mpc.

In the  first part  of this  review, we describe  the universe  -- its
structure,  composition and  global  properties.  Then  we proceed  to
discuss our  current understanding of its origin  and early evolution,
emphasizing the  deep connections between physics on  the smallest and
largest scales.  We end by discussing some recent and more speculative
ideas from  theory, as  well as posing  some of the  ``big questions''
confronting cosmology today.

\section{TAKING THE MEASURE OF THE UNIVERSE}

\subsection{Cosmological Framework}

The framework for  understanding the evolution of the  universe is the
hot    big-bang    model,   technically    referred   to    as   the
Friedmann-Lemaitre-Robertson-Walker  (FLRW) cosmological  model.
Grounded  in Einstein's  theory of  general  relativity, this model
assumes that  on the  largest scales the  universe is  homogeneous and
isotropic, features which have now been confirmed observationally.

The  FLRW  model  incorporating  inflation,  can be  described  by  16
cosmological parameters  that we group  here into two  categories (see
Table I).
The first 10  parameters describe the expansion, the  global geometry, the
age and the composition of  the underlying FLRW model, while the final
6 describe the deviations from exact homogeneity, which at early times
were  small,  but  today  manifest  themselves  in  the  abundance  of
cosmic structure, from galaxies to superclusters.


The Friedmann equation governs the expansion rate and relates  several
of the first 10 parameters:

\begin{eqnarray}
H^2 \equiv \left( {\dot a\over a} \right)^2 & = & { 8\pi G
	\rho_{\rm tot}\over 3} - {1\over R_{\rm curv}^2} \nonumber \\
	\rho_{\rm tot} & = & \Sigma_i \, \rho_i\,, \qquad
	R_{\rm curv}^2  \equiv  k/a^2
\end{eqnarray}

\noindent
where $H$  is the  expansion rate, $a(t)$  is the cosmic  scale factor
(which  describes  the separation  of  galaxies  during the  expansion),
$\rho_{\rm tot}$ is the mass-energy density, and $R_{\rm curv}$ is the
curvature  radius.  The  well known  cosmological redshift  $z$ (which
relates the  observed wavelength  of a photon  $\lambda_{R}$ when
received at  time $t_R$, to  its restframe wavelength  $\lambda_{E}$ when
emitted at time $t_E$) is directly  tied to the change in cosmic scale
factor $a(t)$: $1+z \equiv \lambda_{R}/\lambda_{E} = a(t_R)/a(t_E)$.

{}From the Friedmann equation it follows that the total mass--energy
and spatial curvature $k$ are linked:
\begin{equation}
R_{\rm curv,0}  =  H_0^{-1}/|\Omega_0 - 1|^{1/2}
\end{equation}
where  the subscript  `0' denotes  the  current  value  of the  parameter,
$\Omega_0 \equiv \rho_{\rm tot}/\rho_{\rm crit}$ and $\rho_{\rm crit} \equiv
3H_0^2/8\pi  G$  is  the  so-called ``critical  density''  that  separates
positively  curved  ($k>0$),  high-density universes  from  negatively
curved  ($k<0$), low-density  universes.  Recent  measurements  of the
anisotropy of the cosmic microwave background have provided convincing
evidence that  the spatial  geometry is very  close to  being uncurved
(flat,  $k=0$), with $\Omega_{0}=  1.0 \pm  0.03$ (\onlinecite{deBernardis:2002}).

The currently known components of the Universe include ordinary matter
or baryons ($\Omega_B =\rho_B/\rho_{crit}$), cold dark matter ($\Omega_{\rm
CDM}$),  massive  neutrinos  ($\Omega_{\nu}$),  the  cosmic  microwave
background and other forms of radiation ($\Omega_{\rm rad}$), and dark
energy ($\Omega_{\rm X}$).  The values for these densities are derived
empirically, as discussed  below, and sum, to within  their margins of
error, to the critical density,  $\Omega_0 = 1$, consistent
with the determination of the curvature, $k = 0$.

The  second   set  of  parameters,  which   broadly  characterize  the
individual deviations  from homogeneity, describe (a)  the tiny ($\sim
0.01\%$) primeval  fluctuations in the  matter density as  encoded in
the CMB,  (b) the inhomogeneity  in the distribution of  matter today,
and  (c) the  possible  spectrum of  gravitational  waves produced  by
inflation.  The initial spectrum  of density fluctuations is described
in terms  of its  power spectrum  $P(k)$, which is  the square  of the
Fourier transform  of the  density field, $P(k)  \equiv |\delta_k|^2$,
where  the  wavenumber  $k$  is  related  to  the  wavelength  of  the
fluctuation,  $k=2\pi/\lambda$.   (Galaxies like ours are formed
from perturbations of wavelength $\lambda\sim 1$\,Mpc.)
The  primordial  power  spectrum  is
described  by a power  law, $P(k)  \propto k^n$,  where a  power index
$n=1.0$  corresponds to  fluctuations in  the  gravitational potential
that are the same on all scales $\lambda$ (so-called scale invariant).
The scale-invariant spectrum is predicted by inflation and agrees well
with  current  observations.  The  overall  amplitude  of the  density
perturbations  can be described  by either  $\sqrt{S}$, the
CMB quadrupole anisotropy produced
by the  fluctuations   or $\sigma_8$,
the amplitude of  fluctuations on a scale of  $8h^{-1}\,$Mpc, which is
found from observations to be of order unity.

Accurately  measuring these parameters presents  a significant
challenge.   As   we  now  describe,  thanks  to   major  advances  in
technology,  the challenge  is  being  met, and  in  some cases,  with
independent  measurements  that  check  the consistency  of  both  the
theoretical framework, and the results themselves.

\subsection{The Expansion of the Universe}

The expansion of the universe was discovered in 1929 by Edwin Hubble,
who measured the distances to a sample of nearby galaxies, and
established a correlation between distance and recession velocity.
The slope of this relation is the Hubble constant.  Large systematic
uncertainties in determining distance have made an accurate
determination of the Hubble constant a challenge, and only recently
have improvements in instrumentation, the launch of the Hubble Space
Telescope (HST), and the development of several different measurement
methods led to a convergence on its value.  Accurate distances to
nearby galaxies obtained as part of an HST Key Project have allowed
calibration of 5 different methods for determining the distances to
galaxies out to 500\,Mpc (\onlinecite{Freedman:2001}). All the
techniques show good agreement to within their respective
uncertainties, and yield a value
$$H_0 = 72 \pm 2 \pm 7\,{\rm
km\,sec^{-1}\,Mpc^{-1}},$$
where the error bars represent 1-$\sigma$
statistical and systematic uncertainties, respectively (see
Figure 3).  Because of the importance of its value
to so many cosmological quantities, and because of
its historically large uncertainty, $H_0$ is often written
as $H_0 = 100h\,{\rm km\,sec^{-1}\,Mpc^{-1}}$, so that
$h=0.72\pm 0.02\pm 0.07$.

The largest contributions to these quoted uncertainties
result from those due to the metallicity of Cepheids, the distance to
the Large Magellanic Cloud (the fiducial nearby galaxy to which all
Cepheid distances are measured relative to), and the calibration of
the Wide Field Camera on HST. Other groups using similar techniques
(\onlinecite{Saha:1997}) find a lower value of H$_0$ ($\sim$60
km/sec/Mpc). The reasons for the difference are many, as described
further in \onlinecite{Freedman:2001}, but overall the determinations
are consistent to within the measurement uncertainties.  Recent
measurements of H$_0$ based on two completely independent techniques,
the Sunyaev-Zeldovich method and the measurement of time delays for
gravitational lenses (\onlinecite{Reese:2000, Keeton:2000}), are
yielding values of H$_0$ $\sim$ 60 km/sec/Mpc with systematic errors
currently still at the 20-30\% level.  New results from the WMAP
satellite, discussed in the postscript to this article, give H$_0$ =
71 $\pm$ 4 km/sec/Mpc.

Because light  from very distance  galaxies was emitted long  ago, the
Hubble  diagram also  provides a  means  of probing  the expansion  at
earlier times.   For many decades,  efforts have been  directed toward
measuring what was almost universally  expected to be a slowing of the
expansion over  time due to the  gravity of all  the matter.  However,
observations  by  two  independent  groups have  found  that
supernovae  at high  redshifts  are fainter  than  predicted for a
slowing expansion and indicate that the expansion is actually speeding
up (see  Figure 4) (\onlinecite{Perlmutter:1999,Riess:1998}).
Although systematic effects due to intervening dust or
evolution  of  the supernovae  themselves  could  explain
such a dimming of high-redshift supernovae, several tests
have failed to  turn up any evidence for  such effects.  Apparently,
the universe  is now undergoing  an acceleration, with
the repulsive gravity of some  strange energy form -- dark  energy --  
at  work.  There is weak evidence in the supernova data for
an earlier ($z> 1/2$), decelerating phase (\onlinecite{TurnerRiess:2002}).
Such a decelerating phase is expected on theoretical grounds (more later),
and establishing its existence (or absence!) is an important goal
of future supernova observations.

The  remarkable fact that the expansion is speeding up, rather
than slowing down, can be accounted for  within Einstein's theory,
as the  source of  gravity is
proportional to $(\rho + 3p)$,  where the pressure $p$ and
energy density $\rho$ describe the bulk properties of
the ``substance''.  (For ordinary or even nonbaryonic dark
matter, $p=0$, while for photons
and relativistic particles, $p=\rho /3$.)  A substance
that is very elastic, i.e., with pressure more negative
than one third its energy density, has repulsive gravity
in Einstein's theory (more later).  Of course, it could well
be that the root cause of cosmic acceleration
is not new stuff (i.e., dark energy), but involves a
deeper understanding of gravity.

The deceleration parameter was introduced to quantify the slowing of
the expansion; it is related to the mass--energy content of the Universe:
\begin{equation}
q_0 \equiv {-(\ddot a /a)_0\over  H_0^2} = {\Omega_0\over 2} + {3\over
2} w_X \Omega_X \simeq -0.67\pm 0.25
\end{equation}
where  $w_X  \equiv  p_X/\rho_X$  characterizes the  pressure  of  the
dark-energy component.  ($w_X$ need not be constant; for simplicity,
we shall assume it is.)  In the absence of dark energy, a flat Universe
would decelerate  by its own self-gravity (i.e., $q_0 =0.5$), 
whereas  dark energy allows
for acceleration.  The supernova measurements are consistent with $w_X
=  -1$ and  $\Omega_X  =  0.7$.  Independent  confirmation  of such  a
startling result  is extremely important.  As  discussed below, strong
indirect  evidence for  an additional  energy component  comes  from a
comparison of the density  of matter with measurements of $\Omega_{0}$
from fluctuations in the CMB.

Dark   energy, a ``mysterious substance'' whose pressure is
negative and comparable in magntiude to its energy density,
apparently   accounts  for  two-thirds   of  the
matter--energy budget  of the universe  and has no  clear explanation.
Understanding its  nature presents one  of the greatest  challenges in
both cosmology and particle physics.

\subsection{The Matter Composition of the Universe}

While we  know more about the  other one-third of the  universe -- the
matter part --  important questions remain.  According to  the current best
census, the  visible part of  ordinary matter -- that  associated with
stars -- contributes only about 1\% of the total. What we can see with
telescopes is literally the tip of an enormous iceberg.

The rest of  the matter in the universe is dark,  and its existence is
inferred  from its  gravitational effects.   While the  case  for dark
matter holding together galaxies (as well as clusters of galaxies) has
been around  for a long time (\onlinecite{Zwicky:1933,Rubin:1980}),  the nature of
the dark  matter in the  universe is still  unknown. In fact, we still speak
with more certainty about what dark  matter is known not to be.  Based
upon simple accounting, we have  all but eliminated the possibility of
dark  matter being  made of  neutrons and  protons, and  established a
strong case for a new form of matter.

The accounting  of ordinary  matter involves three  different methods,
all  of which  arrive at  the same  answer.  The  most  precise of these
methods comes  from consideration  of  the  formation of  light
elements  during big-bang  nucleosynthesis (BBN).  Hydrogen,  helium, deuterium,
and  lithium  are  produced  in  the  first few  minutes  of  the  big
bang. However,  only if  the density of  ordinary baryons is  within a
narrow  range is  the  predicted production  consistent  with what  we
actually  measure (see Figure 5).   The production  of deuterium  is the
most  sensitive indicator  of the  baryon density.   Measurements made
with  the 10-meter  Keck  Telescopes  of the  amount  of deuterium  in
high-redshift clouds  of gas (seen  by their absorption of  light from
even  more distant  quasars in the Lyman series of lines),
together  with  the theory  of big-bang
nucleosynthesis  yield a  density of  ordinary matter  of  $3.8\pm 0.2
\times 10^{-31}\,{\rm g\, cm^{-3}}$, or  only about 4\% of the critical
density (\onlinecite{Burles:2001}).

Two  other  determinations  are  consistent with  the  nucleosynthesis
argument: First, the net absorption of light emitted from very distant
quasars by  intervening gas (which exists in clouds of gas
known as the Lyman-alpha forest after the multiplicity of redshifted
absorption features produced by the individual clouds)
indicates  a similar value for  the baryon
density. This  probes ordinary matter at a time and  place when the
bulk of  the baryons are  still expected to  still be in  gaseous form
($z\sim 3-4$).  The second constraint comes from measurements of
the CMB, which yield an independent baryon density consistent
with that determined from nucleosynthesis.
Our best accounting of ordinary matter comes from
this early, simpler time, before  many stars had yet formed.

Our accounting of baryons at the present epoch, in the local universe,
is  not  as  complete.   Baryons  in  stars  account  for  only  about
one-quarter of all the baryons;  the rest are optically dark.  While a
number of possibilities for the  baryonic dark matter (from planets to
black  holes) have  been  considered,  it now  appears  that the  most
plausible reservoir  for most  of the unseen  baryons is warm  and hot
ionized gas surrounding galaxies within groups and clusters.  In fact,
in rich clusters the amount  of matter in hot intercluster gas exceeds
that in  stars by  a large factor.   But since  only a few  percent of
galaxies are found  in these unusually rich clusters,  the bulk of the
dark baryons are still unaccounted for.

While not  all of  the dark baryons  are accounted for,  baryonic dark
matter itself  only accounts for  about one-tenth of all  dark matter.
The evidence that  the total of amount of dark  matter is much greater
-- about  one-third of the  critical density  -- has  gradually become
firm,  as  several, independent  (and  increasingly higher  precision)
measures have yielded concordant results (\onlinecite{Sadoulet:1999,Griest:2000}).

Clusters of  galaxies provide a laboratory for  studying and measuring
dark  matter in  a variety  of ways.   Perhaps most  graphically, dark
matter can be  seen in its effect on  more distant background galaxies
whose  images   can  be   distorted  and  multiplied   by  dark-matter
gravitational lensing effects.  This  and other techniques (applied in
the x-ray, radio, and optical)  have determined the ratio of the total
cluster  mass  to ordinary  matter  (predominantly  in  the hot  x-ray
emitting intracluster gas): averaged over more than fifty clusters the
ratio is  about 8 (\onlinecite{Mohr:1999,Grego:2001}).  Assuming that clusters  provide a ``representative
sample'' of matter in the Universe,  the total amount of matter can be
inferred from  the baryon density. That  number is about  one third of
the critical density.

What then is this nonbaryonic  dark matter?  The working hypothesis is
weakly  interacting   elementary  particles  produced   in  the  early
universe.   Before  discussing  specific  particle  candidates,  let's
review  the  constraints   from  astrophysical  observations.   First,
because dark matter is  diffusely distributed in extended halos around
individual galaxies or  in a sea through which  cluster galaxies move,
dark-matter particles must not interact with ordinary matter very much,
if at all.  Otherwise, dark matter would by now have dissipated energy
and relaxed to the more concentrated structures where only baryons are 
found.  At the very least, we can be confident that the constituents of
nonbaryonic dark matter are uncharged, and have only very weak interactions.

In addition, the formation of  structure in the universe tells us that
early on  dark-matter particles must  have been cold (i.e.,  moving at
non-relativistic    speeds)   rather    than    hot   (i.e.,    moving
relativistically).   If  the dark  matter  had  been  hot, then  these
fast--moving  particles would  have smoothed  out the  smaller density
irregularities, which seed the  formation of galaxies and clusters, by
streaming from high-density regions to low-density regions.  The first
objects  to   form  would  have  been  the   largest  structures  (the
superclusters) and  smaller objects (galaxies) would  have only formed
later   by  fragmentation.    However,  this   is   inconsistent  with
observations.

The deep  image of the  sky obtained by  HST in 1995 (the  Hubble Deep
Field; see Figure 6),  along  with other  observations  by
ground-based  telescopes,
identified the epoch when most  galaxies formed as a few billion years
after the big bang (at redshifts of  order 1 to 3).  The Sloan Digital Sky
Survey,  as   well  as  x-ray   observations  from  space   and  other
ground--based  telescopes,   have  shown  that   clusters  form  later
(redshifts  less than  about  1).  Finally,  superclusters, which  are
loosely  bound  collections  of  a  few  clusters,  are  forming  just
today. This sequence is inconsistent with hot dark matter.

Nonetheless, there is at least one hot dark-matter particle that we do
know  exists -- the  neutrino. Two  experiments, one
undertaken in  Canada, the other  in Japan, now provide  evidence that
neutrinos have  mass (\onlinecite{Fukuda:2002,Ahmad:2001,Ahmad:2002}).  The  experiments, which are
studying  solar neutrinos  and  atmospheric neutrinos,  have placed  a
lower  limit   on  the  mass   of  the  heaviest  neutrino   at  about
0.05\,eV. This implies that neutrinos contribute at least 0.1\% of the
mass--energy  budget  of  the  universe.   However,  the  cosmological
considerations just discussed cap  the contribution of neutrinos -- or
any hot  dark matter  candidate --  to be less  than about  5\%.  This
leaves the  bulk of the dark  matter still to be  identified.  We will
return to the other particle candidates for dark matter later.

\subsection{The Cosmic Microwave Background}

Today, CMB  photons, while  very numerous (there  are about  2 billion
photons for every hydrogen atom)  account for a negligible fraction of
the mass--energy  budget (about 0.01\%).   Still, they play a central
role in cosmology.    First, at early times,  the CMB was the
dominant part of the mass--energy budget, from which we ascertain that
the infant  Universe was a  hot thermal bath of  elementary particles.
Second, photons from the CMB  interacted closely with matter until the
temperature of the  Universe had cooled enough for  the ionized plasma
to  combine and  form neutral  atoms, allowing  the photons  to stream
past.  At  this ``last-scattering surface''  of the CMB,  the Universe
was about 400,000\,years old, and  about 1100 times smaller than it is
today. The  CMB is a  ``snapshot'' of the  Universe at a  much simpler
time.

The  CMB  measurements  are a  striking  example  of  a new  level  of
precision  now being  made  in cosmology.   NASA's  COBE satellite,  a
four-year mission  launched in 1989,  measured the temperature  of the
background radiation  to better  than one part  in a thousand,  $T_0 =
2.725\pm  0.001  \,$K  (\onlinecite{Mather:1999}),  and discovered  tiny  (tens  of
microKelvin) variations in the temperature  of the CMB across the sky.
These  tiny   fluctuations  arise  from  primeval   lumpiness  in  the
distribution of matter.  In  the early Universe, outward pressure from
the CMB photons, acting counter to  the inward force of gravity due to
matter, set  up oscillations whose frequencies are  now seen imprinted
in the CMB fluctuations.   Evidence of these ``acoustic oscillations''
can   be  seen   when  the   fluctuations  are   described   by  their
spherical-harmonic  power spectrum  (see Figures 7-9).   In late
2002,  the DASI Colloboration detected  the last  feature
predicted for  the CMB: polarization (\onlinecite{Kovac:2002}).
Because the CMB radiation is not isotropic (as evidenced by the
anisotropy seen across the microwave sky) and Thomson scattering off
electrons is not isotropic, CMB anisotropy should develop about a
5\% polarization.

The  precise shape  of the  angular  power spectrum  of anistropy  and
polarization  depends in  varying  degrees upon  all the  cosmological
parameters  in Table  I, and  so CMB  anisotropy encodes  a  wealth of
information  about the  Universe.   With a  host  of ground-based  and
balloon-borne  CMB experiments  following COBE,  a NASA  space mission
(the Microwave Anisotropy Probe, MAP) now taking new data, and with an
European Space Agency (ESA)  mission planned  for
launch in  2007,  we are  in  the midst  of
realizing  the  potential  of  the  CMB as  a  probe  of  cosmological
parameters.  A  summary of the progress includes  determination of the
curvature, $\Omega_0 = 1.03 \pm  0.03$, the power law index of density
perturbations, $n=1.05\pm 0.09$, the  baryon density $\rho_B = 4.0 \pm
0.6 \times 10^{-31}\,{\rm g\,cm^{-3}}$, and the matter density $\rho_M
= 2.7\pm 0.4 \times 10^{-30}\,{\rm g\,cm^{-3}}$.  The uncertainties in
all of these quantities are expected  to diminish by at least a factor
of ten.

As mentioned above, the CMB value for the baryon density is consistent
with that determined from BBN.  This not only provides confidence that
ordinary matter accounts  for a small fraction of  the total amount of
matter,  but also  is  a  remarkable consistency  test  of the  entire
framework.  The CMB provides independent, corroborating evidence for a significant
component  of dark energy  through the  discrepancy between  the total
amount of matter and energy (critical density) and that in matter (1/3
of  the  critical density).   Finally,  the  measurements  of the  CMB
multipole spectrum  are consistent with the emerging  new cosmology: a
flat Universe with dark  matter and dark energy.

Establishing a  reliable accounting of the matter  and energy in  the Universe (see
Figure 10) is a major achievement; but,  we still have much more to learn
about  each component and  almost everything  to understand  about the
``strange recipe.''   Moreover, because the energy  density of matter,
photons  and  dark energy  each  change  in  distinctive ways  as  the
universe expands, the mix we see today must have been different in the
past and will be different in the future.

The energy per photon (or  per relativistic particle) is redshifted by
the  expansion (decreasing  as  $a^{-1}$) and  the  number density  of
photons is diluted by the  increase in volume (as $a^{-3}$), resulting
in a  total decrease in  the energy density proportional  to $a^{-4}$.
The energy density in matter is  diluted by the volume increase of the
universe, so  that it  decreases as $a^{-3}$.   The energy  density in
dark energy  changes little (or not  at all) as  the universe evolves.
This means that the  Universe began with photons (and other forms
of  radiation)  dominating the  energy  density  at  early times  ($t<
10^4\,$yrs),  followed by  an era  where matter  dominated  the energy
density, culminating  in the present  accelerating epoch characterized
by a transition to a universe dominated by dark energy.

\subsection{The Structure of the Universe Today}

The distribution of galaxies in the local universe reveals a striking,
hierarchical pattern with  a variety of forms such  as galaxy clusters
and  superclusters,  voids  and  bubbles, sheets  and  filaments  (see
Figure 1).  In  the past 20 years,  the volume of  the universe surveyed
has  grown  immensely, particularly  with  the  recent development  of
multi-fiber and multi-slit spectrographs,  which allow redshifts to be
measured for  hundreds of  galaxies at one  time.  In  the mid-1980's,
redshifts for about 30,000  galaxies were measured individually with
velocities of up to 15,000 km/sec as part of the CfA survey (\onlinecite{GellerHuchra:1989}).

Unexpected, large-scale  structures (walls and bubbles) were
revealed with sizes that continued  to grow as the survey volumes
expanded.  In the mid-1990's, about 26,000 additional galaxy redshifts
with velocities up to 60,000 km/sec were measured with
a multi-fiber spectrograph as part
of the  Las Campanas survey.   The larger (but more  sparsely sampled)
Las Campanas survey  found no new larger structures:  the universe had
finally  revealed its  homogeneous nature  on the  largest  scales, as
expected  from  the  uniformity   of  the  CMB.   The  most  ambitious
large-scale-structure  surveys   to  date  are   the  Anglo-Australian
Two-degree Field  Galaxy Redshift Survey (2dFGRS),  which has compiled
almost  250,000  redshifts covering  about 5\%  of  the  sky, and  the
on-going Sloan Digital Sky Survey  (SDSS), which now has close to half
of the 600,000 galaxy redshifts  it plans to obtain over about 25\% of
the sky.

The  simplest  description  of  galaxy clustering  is  the  two--point
correlation  function,  which  measures  the excess  probability  over
random of finding  two galaxies separated by a  given distance.  It is
found empirically to follow a simple power law,
$$\xi = (r/6h^{-1}\,{\rm  Mpc})^{-1.8}\,,$$ which implies that finding
another galaxy within  $6h^{-1}\,$Mpc from a given galaxy  is twice as
likely as finding a  galaxy within a circle  of radius 6~Mpc  placed randomly on
the sky.   The Fourier  transform of the  correlation function  is the
previously discussed  power spectrum  of the distribution  of galaxies.   
The power spectrum can be  directly compared with
theoretical  predictions  from  inflation  and cold  dark  matter.   A
complication  in this  comparison is  the  extent to  which the  light
observed in  galaxies faithfully traces the distribution  of mass.  It
is now  known that galaxies are  slightly (10\% or  so) more clustered
than the mass,  and that this ``biasing'' is  most pronounced on small
scales.   That being said,  the observed  and predicted  power spectra
(shown in  Figure 11)  compare well.

On the largest scales, the power spectrum, which measures the level of
inhomogeneity  today, can also  be compared  with measurements  of the
anisotropy of the  CMB. This measures the level  of inhomogeneity when
the Universe was only 400,000 years old and the structure existed only
as  the  seed fluctuations.   Because  the  growth of  inhomogeneities
depends  upon the  composition of  the Universe,  the  comparison with
theory depends also upon cosmological parameters.  When the comparison
is made, there is reassuring consistency.

\subsection{The Age of the Universe}

The time back to the big bang depends upon $H_0$ and the expansion history,
which itself depends upon the composition:
$$ t_0  = \int_0^\infty  {dz \over (1+z)H(z)}  = H_0^{-1}\int_0^\infty
{dz\over  (1+z)[\Omega_M(1+z)^3 +\Omega_X  (1+z)^{3(1+w_X)}  ]^{1/2} }
\,,$$where $\Omega_M = \Omega_{\rm CDM} +\Omega_B+\Omega_\nu$
is the total mass density.

For  a   universe  with   a  Hubble   constant   of  $72\,{\rm
km\,sec^{-1}\, Mpc^{-1}}$ and matter  contributing 1/3 and dark energy
2/3 to the overall mass-energy density,  the time back to the big bang
is  13\,Gyr.  Taking  account of  the uncertainties  in $H_0$  and the
composition, the uncertainty  in the age of the  universe is estimated
to be about $\pm 1.5$\,Gyr.  The expansion age can also be determined
from CMB  anisotropy, but  without recourse to  $H_0$, and it  gives a
consistent age, $t_0 = 14\pm 0.5\,$Gyr (\onlinecite{Knox:2001}).

The expansion age can  also be  checked for  consistency against  other cosmic
clocks.   For example, the  best estimates  of the  age of  the oldest
stars in the universe are obtained  from systems of $10^5$ or so stars
known  as globular  clusters.   Stars spend  most  of their  lifetimes
undergoing nuclear  burning of hydrogen  into helium in  their central
cores.   Detailed  computer models  of  stellar  evolution matched  to
observations  of globular-cluster stars yield  ages  of about  12.5
billion years,  with an uncertainty  of about 1.5\,Gyr (\onlinecite{KraussChaboyer:2002}).
These estimates  are also  in good agreement  with other  methods that
independently measure the  rates of cooling of the  oldest white dwarf
stars, and techniques that  use various radioactive elements as cosmic
chronometers (\onlinecite{Oswald:1996}).  Finally, with the assumption
that $w_X=-1$, the type Ia supernova data can constrain the
product of the age and Hubble constant independent of either quantity,
$H_0t_0 = 0.96\pm 0.04$ (\onlinecite{Tonryetal:2003}).  This is
consistent with the product of the two, $(H_0 = 72\pm 8\,{\rm km\,
sec^{-1}\,Mpc^{-1}})\times (t_0 = 13\pm 1.5\,{\rm Gyr}) = 0.96 \pm
0.16$.

In summary, all the ages  for the universe are  consistent with a  
consensus age of  about $13\pm 1.5\,$Gyr.

\subsection{Recap: The New Emerging Cosmology}

Using  a host  of diverse  observations made  possible by  advances in
technology, we have now defined the basic features of the Universe: it
is  spatially  flat and  13\,Gyr  old  with  a currently  accelerating
expansion,  comprised of  one third  dark matter  and two  thirds dark
energy.  The variety and  redundancy of observations is important: the
overall picture does  not depend on just a  single measurement or type
of measurement.  Consistency checks  on the age,  the baryon and matter
densities, and the total density and spatial  curvature have  increased
our confidence  both in the description as well as  in  the  overall
framework.   We now move from discussion
of the basic  features of the universe to  the underlying physics that
forms the basis for our current understanding.

\section{UNDERSTANDING THE UNIVERSE}

In addition to breakthrough  observations, creative new ideas are also
driving progress  in cosmology.   Not only do  we know more  about the
universe, but our  understanding is deeper, and the  questions that we
are asking are more profound.   Still, our understanding of the origin
and evolution of the universe has  not yet caught up with what we know
about it.   But a  vital ingredient  in  furthering our
understanding and  shaping our questions  has been the  recognition of
the  connections that exist  between the  elementary particles  on the
smallest scales, and the universe on the largest.

\subsection{Origin of Structure}

The  abundance of structure  that has  been mapped  out today  -- from
galaxies of  mass as small  as $10^6M_\odot$ to superclusters  of mass
exceeding  $10^{16}M_\odot$ -- speaks  to a  remarkable transformation
that  occurred in  the early  universe as small primeval
inhomogeneities  in the distribution
of matter were amplified by the attractive forces of gravity
($1 M_\odot \simeq 2\times 10^{33}\,$g refers to the mass of the sun).

During the earliest moments the primordial fluctuations did not grow because
the  expansion was  controlled by  radiation, and  radiation  does not
clump.  When the universe became matter-dominated, the inhomogeneity
then  grew at  precisely the  same rate  as the  cosmic  scale factor,
($\delta  \rho /\rho  ) \  \propto\ a(t)$.   Because the  size  of the
universe grew by a factor of about 10,000 during the matter dominated
era, initial  fluctuations of amplitude 0.01\%  or so are  all that is
needed  to seed  the  nonlinear structure  seen  today.  Formation  of
structure then  ceased a  few billion years  ago when  the accelerated
expansion began, pushing the existing structures apart.

The required primordial lumpiness should have  left its signature on the CMB in
the form  of temperature fluctuations of order  ten microKelvins.  And
this  is precisely  what  has  been measured,  both  in amplitude  and
variation with angular  scale (cf. Figure 9) .  Being  able to account for
how the structure seen today arose from a nearly homogeneous beginning
is a major success of the FLRW model.

Of  course, a  natural  question arises:  What  is the  origin of  the
primeval lumpiness?   Cosmologists have a working model  for this: the
seeds  of  structure  arose  from  the stretching  and conversion
of  quantum  noise originating on  subatomic length scales  to 
density inhomogeneity on astrophysical
length scales, caused by a tremendous  burst of expansion called inflation.
Inflation, coupled with the  idea of including nonbaryonic dark matter
in the universe, has led to a predictive and descriptive theory of how
the  structure  in the  universe  plausibly  arose.  This  descriptive
theory is known as Cold Dark Matter (CDM) (\onlinecite{Blumenthal:1984}).  From the
simple  starting  point of  cold  dark  matter and  inflation-produced
lumpiness  follows a highly  successful picture  for the  formation of
structure in the universe.

The  defining prediction  of  CDM  is that  structure  forms from  the
``bottom up'': first galaxies,  then clusters of galaxies, and finally
superclusters.  This ordering, now confirmed  by observations, follows
from  the fact  that  the degree  of  lumpiness is  larger on  smaller
scales, so that the smaller objects become gravitationally bound, stop
expanding, and collapse back on themselves first.

As computing power and numerical techniques have improved, simulations
of the evolution of structure in the universe have become increasingly
sophisticated,  providing  additional insight  into  the evolution  of
cosmic structure.   The initial ingredients for  these simulations are
specified by the CDM paradigm, and the ensuing growth of structure via
the force of gravity is  now computed by  following the motions  of more
than a  billion particles.  The  properties of the  simulated universe
(correlation  function, power  spectra, and  masses and  abundances of
galaxies, etc)  are calculated for different recipes  (dark matter and
baryon densities,  with and without dark energy,  and different values
of  the  cosmological  parameters)   and  can  all  be  compared  with
observations.     Numerical   simulations    were    instrumental   in
demonstrating the  failure of structure  formation in the  presence of
hot dark matter,  and the success of flat CDM  models with dark energy
in matching  the observed distribution  of structure seen  today.  The
successful predictions of CDM  go beyond being merely descriptive, and
now include many quantitative predictions.

That being said, the successes of CDM largely  involve predictions that follow from the
basic   paradigm  and   the   action  of   gravity  alone.    However,
understanding the development of  the structure we see with telescopes
requires an understanding of how  baryons form into stars in galaxies.
Complicated gas  dynamics come into play and  there may even be
feedback  from  the energetics of star formation on the dark-matter
structures themselves. Crisp predictions become more  difficult.
Understanding how real galaxies form and evolve is a rich subject with
many   outstanding   questions  whose   answers   will  require   more
observations  as well  as detailed  astrophysical modelling  that goes
beyond the gravity of cold dark matter alone.  Pinning down the basic
cosmological framework has helped significantly by removing uncertainty
associated with the evolution of the Universe.

\subsection{The Expansion}

Hubble's discovery of the expansion of the universe changed our cosmic
perspective  forever  -- we  now  know that  we  live  in an  evolving
universe  with  a   big-bang  beginning.   Its  interpretation  within
Einstein's  theory  of gravity  was  a  striking  confirmation of  the
dynamical nature  of space  and time  -- the expansion  is due  to the
stretching of space itself.   However, within the framework of general
relativity  there is  no answer  to the  most basic  question  -- what
happened  just  before  the  big  bang to  get  the  expansion  going?
According  to  general  relativity,  the  big bang  was  the  singular
creation of matter, energy,  space and time itself.  If correct,
then there was  no ``before'' before the big bang --  a neat and tidy,
if not entirely satisfying, answer.

Since Einstein's theory does  not incorporate quantum mechanics, there
is reason to believe that it  is not complete or fully applicable
around  the time  of the  big bang.   Answering the  ``before  the big
bang''  question  most  likely  still depends  upon  marrying  quantum
mechanics to gravity, and applying  that merged theory to the earliest
moments of  the universe.  Currently,  superstring theory is  the most
promising idea for making such  a connection, and we will return later
to some of the speculations based upon it.

Even within  general relativity  there are conceptual  questions about
the beginning.  Not all big-bang models necessarily lead to a universe
like  ours.   Unless the  initial  conditions  were  ``just so''
(see, e.g., \onlinecite{Hogan:2000}),  the
universe might well  have recollapsed long ago; or  it might have gone
into a coasting phase of indeterminate duration; or the expansion need
not necessarily have  been the same in all  directions (isotropic), or
there might  not necessarily have  been large-scale regularity  in the
distribution of  matter (homogeneity), as  we observe in  the universe
today.

Apparently a  special beginning  is required.  There  are two  ways to
read  this:  The  first  is  that  the state  of  the  universe  today
accurately pins down the initial conditions, a point of view advocated
by a  few (\onlinecite{Penrose:1979}).   The second  is to look  for a way  to get
around the special  conditions, a road that led to  the idea of cosmic
inflation.

In 1980, Alan  Guth pointed out that an  early, brief period of
very   rapid  (indeed  exponential)  expansion,   now  called
inflation, could change the cosmic landscape dramatically (\onlinecite{Guth:1980}).
He  had in  mind a  scenario  wherein the  universe got trapped in a
``false-vacuum   state''  during   a  cosmological   phase  transition
associated with the breaking of  symmetry between the strong, weak and
electromagnetic forces.  Although this  specific idea does not seem to
work,   it nevertheless led  to  a paradigm  for  inflation  based on  the
potential  energy  associated with  an  as-of-yet hypothetical  scalar
field called the inflaton (\onlinecite{Linde:1983,AlbrechtSteinhardt:1982}). (If it exists, this field is
distantly related  to the Higgs scalar field  that particle physicists
believe explains why particles have mass.)

According to inflation,  small bits of the universe  underwent a burst
of  expansion when  the universe  was extremely  young  ($t\ll 10^{-2}
$sec).  This  expansion flattened the  local geometry in the  same way
inflating a balloon makes a  fixed region look flatter and smoother as
the balloon inflates (in the case of the universe the blow-up factor exceeded
a factor of $10^{40}$!).  The conversion of the scalar-field potential
energy  into particles and  photons accounts  for the  tremendous heat
content of  the big bang and initiates  the early, radiation-dominated
era.   During   the  conversion   of  scalar  field   energy,  quantum
fluctuations in  the inflaton field on subatomic  scales, stretched to
astrophysical  size  by  the  tremendous expansion,  became  the  seed
density inhomogeneities.  Moreover, quantum fluctuations in the metric
of spacetime  itself lead to  a predicted spectrum  of long-wavelength
gravitational waves.

Thus, cosmic  inflation explains the  smoothness of the  universe, the
heat  of  the  big  bang,   and  it  predicts  a  flat  universe  with
characteristic seed  irregularities in  the matter distribution  and a
spectrum of  gravity waves.  It  also says that  all we can  see arose
from ``our  big bang,'' one  of many bursts of  inflationary expansion
that took place early on.

As  of  yet  there  is  no single,  agreed-upon  model  of  inflation.
However, all  models are based  upon the paradigm described  above and
make three  firm predictions:  (i) a spatially  flat universe;  (ii) a
nearly  scale-invariant  distribution  of density  perturbations;  and
(iii)  a  nearly  scale-invariant  spectrum  of  gravitational  waves.

In the early  1980s when inflation was gaining  sway with cosmological
theorists,  its  prediction  of  a  flat universe  looked  to  be  its
downfall.   The best  measurements indicated  that  matter contributed
only about  10\% of the  critical density, although  the uncertainties
were  quite large  given  that the  amount  of dark  matter was  still
largely  undefined.   But  as  we  saw  above,  recent  determinations
indicate that the universe is  almost certainly flat, with dark matter
contributing one  third of  the critical density  and dark  energy the
other two thirds.

The seed inhomogeneities of inflation were impressed upon the universe
very  early on.   This results  in a  kind of  synchronized  motion of
irregularities   on  different   length   scales,  and   leads  to   a
characteristic  pattern   of   ``acoustic   peaks''  in  the
multipole power spectrum of the CMB anisotropy.  Measurements now show
a clear pattern of acoustic  peaks,  cf.  Figure 9.
Their  relative  heights  also
indicate that the seed  perturbations are consistent with being almost
scale-invariant, again as inflation predicts.

Inflation  has passed its  first  round  of tests.  Over  the next  decade
additional observations, especially  those coming from measurements of
the CMB, will  test inflation more
decisively  and  may  shed  light  on how the inflaton
field  fits into the larger
picture of particle physics.

The  spectrum of  gravitational  waves extending  from wavelengths  of
kilometers to billions  of light years are certainly  beyond the reach
of the  current generation of Earth-based  gravity-wave detectors.
There is some
hope that  gravity-wave-induced polarization signatures  in the cosmic
microwave background  can be detected  with a new generation of
experiments.  If so, not only  would the third prediction of inflation
be  tested,  but  also the  time  when  inflation  took place  would be
identified.

While  inflation is  making  a good  case  for being  included in  the
standard  cosmological model,  it  does not address the biggest question,
how the Universe began.   Superstring
theory shows promise of shedding more light on how the Universe began.
With superstring's  prediction that there are more  than three spatial
dimensions, it  opens new  dimensions in cosmology,  both figuratively
and  literally.   While there  have  yet to  be  successes  or  even
compelling  predictions,  superstring  theory  has  evoked  intriguing
cosmological ideas.  The ``brane  world'' idea holds that our universe
is actually a three-dimensional ``sheet'' in a much higher dimensional
space where gravity exists in the full spacetime, but other forces and
particles are  confined to only three dimensions.   From this paradigm
has come the  idea that the big bang might have  been the collision of
two  such  sheets,  and  even  further  that  such  collisions  happen
repeatedly, adding new life  to the old oscillating universe scenario.
Such ideas are not yet well developed enough to be testable, but their
creativity  speaks to  the vitality  of cosmology.   And it  should be
remembered that  about twenty years  ago inflation seemed  radical and
untestable!

\subsection{The Composition of the universe}

Thirty  years ago  the  universe seemed much  simpler.  We only had
knowledge of ordinary matter, and  even the fact
that  most of  the ordinary matter  did not  reside in  stars was  still  to be
discovered.  Today, we have  a much more complete (and correspondingly
much  more  complicated) accounting,  with  five components:  ordinary
matter, massive neutrinos, cold dark matter, dark energy, and photons,
cf Figure 10. And  now even the ``ordinary'' matter is  not simple -- the
bulk  of  it is  dark,  and in  a  form  yet to  be  firmly identified.

The  leading candidates for  the CDM  particle are  the axion  and the
neutralino,  two hypothetical  elementary particles.   If  they exist,
these particles  would have been  produced in the earliest  moments of
the universe,  and survived in  sufficient numbers to account  for the
dark matter.  Both are new  forms of stable matter, predicted to exist
by theories that attempt to unify the forces of Nature.  But they have
wildly different  masses: a  trillion times smaller  than that  of the
electron for  the axion, and a  hundred times larger than  the mass of
the proton for the neutralino.

If the  cold dark matter hypothesis  is correct, then the  halo of our
own Milky Way should be awash  in axions or neutralinos (or some other
slowly  moving  particle).   While  the  interactions  of  axions  and
neutralinos with ordinary  matter are very weak (and  can be neglected
in almost  all circumstances) specialized detectors  have already been
built  to confirm  (or rule  out)  their presence.   In addition,  the
neutralino,  the lightest  of a  new class  of particles  predicted by
superstring theory, has two other signatures of its existence.  It can
be  produced  at  a  particle accelerator,  given  sufficient  energy.
Alternatively, high-energy  neutrinos produced by  the annihilation of
the few neutralinos that are captured by the Sun could be detected; or
positrons and/or  gamma rays  produced by neutralinos  annihilating in
the halo  might be  found.  With the  efforts underway,  evidence that
axions  or  neutralinos  comprise   the  cold  dark  matter  could  be
forthcoming in the next decade (\onlinecite{Sadoulet:1999,Griest:2000}).

Our state  of understanding  of the origin  of the  various components
comprising the universe varies  widely.  If inflation is correct, then
photons arose  from the  decay of the  potential energy of  the scalar
field that drove inflation.   The existence of quark-based matter that
we  are made  of involves  three elements:  the action  of microscopic
forces that  do not conserve  the net number of  quarks (baryon-number
violation) and break the  symmetry between particles and antiparticles
(referred  to   as  CP  violation),  and  a   departure  from  thermal
equilibrium.   These three  conditions, first  spelled out  by Russian
dissident and physicist Andrei Sakharov in 1967, are necessary for the
universe to develop a slight excess of baryons over antibaryons.  When
the universe  was around $10^{-5}\,$sec  old, the bulk of  the baryons
and  antibaryons annihilated,  leaving the few  residual baryons  for
every 10 billion photons that  now constitute the ordinary matter we see
today.   The details  of ``baryogenesis'' -- which may even involve neutrinos --
are not  fully understood.  Critical in this regard is a better
understanding of  CP violation and neutrinos.

In some ways the emergence of  neutrinos and cold dark matter from the
post--inflation  thermal bath  is  better understood.   At very  early
times,  when temperatures  were  extremely high,  a  kind of  particle
democracy existed,  with roughly equal numbers of  all particle types.
As the temperature dropped below the point where a given species could
be still be produced in pairs  (i.e., where the thermal energy $kT$ is
less than its rest mass energy $mc^2$), the numbers of those particles and
antiparticles decreased rapidly  through mutual annihilation.  Because
of  their small masses  and their  small annihilation  cross sections,
neutrinos never  annihilate, leaving them as abundant today as
CMB photons.  Once  their masses  are known,
their contribution to the mass density follows directly.  From what we
know about neutrino masses, their  contribution is at most comparable to
that of ordinary matter, too small to account for the bulk of the dark
matter.  Nonetheless, neutrinos validate the basic idea of dark matter
in the form of something other than baryons and do play a small role
in the formation of structure.

The  neutralino  story is  more  complicated.   Cosmic neutralinos  do
annihilate significantly; however, once their abundance falls to about
1  per  billion photons,  they  become  so  rarified that  they  cease
annihilating.  Their relic abundance, determined by their annihilation
cross section, turns  out to be in the right range  to account for the
dark matter.

Dark  energy is  the  largest  and most  perplexing  component of  the
universe.  The simplest possibility,  that it is the energy associated
with  quantum   vacuum  fluctuations,  suffers  from   the  fact  that
calculating  how  much   ``quantum  nothingness  weighs''  has  eluded
theorists for more than five  decades and naive estimates are at leats 55
orders-of-magnitude too large!  This  suggests to some that ultimately
it  will be shown  that ``even  quantum nothingness  weighs nothing,''
because that outcome seems more  likely than finding a means to reduce
the naive estimate  by precisely 55 orders-of-magnitude or more.  Yet, if the
dark  energy is  not quantum  vacuum energy,  what is  it?  A  host of
possibilities have  been suggested, from  a mild version  of inflation
involving  an extremely  light scalar  field to  the influence  of new
physics  occurring in the  extra spatial  dimensions, as  predicted by
superstrings. However, the  small magnitude of the dark  energy is not
the only problem.   Another is trying to understand  why at this point
in  time, the  dark  energy is  only  just beginning  to dominate  the
expansion.  It  seems an odd coincidence,  and one that  so far defies
explanation.

Finally, what  can we say about  the destiny of the  universe?  In the
simple universe containing only matter, the destiny of the universe is
linked  to  its geometry:  uncurved  and  negatively curved  universes
expand forever; and positively  curved universes recollapse.  While we
have  determined that  we live  in an  uncurved universe,  adding dark
energy  to the  mix  severs  the link  between  geometry and  destiny.
Depending  upon the nature  of the  dark energy,  a flat  universe can
continue accelerating  forever (if the  dark energy is  quantum vacuum
energy), or  it can slow  down or even  recollapse if the  dark energy
dissipates with time (\onlinecite{KraussTurner:1999}).

\section{LOOKING FORWARD}

In the last two decades, a set of interesting ideas
based upon  unexpected connections between  the quarks and  the cosmos
and the emergence of a new generation of observations and experiments
have transformed cosmology into  a full-fledged, precision  science.  
The ten-microKelvin
fluctuations  in   the  cosmic  microwave   background  radiation  are
constraining  cosmological  parameters   and  shedding  light  on  the
earliest  moments  of  the  universe.   Maps of  the  distribution  of
galaxies and  clusters in volumes approaching  billions of megaparsecs
on  a side are  testing the  cold dark  matter paradigm.   The current
expansion rate  of the universe has  finally been pinned  down to 10\%
precision, and measurements of the  past rate have revealed we are now
in  a period  of cosmic  acceleration.  The  contribution  of ordinary
matter to the overall mass--energy  budget has been shown to be small,
with more  than 95\%  of the universe  existing in new  and unidentified
forms of matter and energy.

Many, creative  theoretical ideas have  emerged that provide a  way to
understand  the expansion of  the universe,  its composition,  and the
origin  of structure.   Still, big  questions remain.   Why  are there
three different  forms of matter/energy  of comparable abundance,
with the transition to   accelerated expansion
occurring  very recently?  How much of  the truth does inflation hold about the
early universe and what  is the  hypothetical inflaton  field that
drove inflation?  What is the  dark matter and the strange dark energy?
Could the  complicated recipe and accelerated  expansion indicate that
we don't yet fully understand gravity?

Astronomers and physicists are in  the midst of carrying out ambitious
new  experiments,  completing  large  surveys  of  the  Universe,  and
commissioning  powerful  new  telescopes  with  novel  technology  and
advanced  instrumentation.   There is  great  promise of  increasingly
sharper tests  of inflation,  cold dark matter,  and dark  energy, and
always the  potential for  further new surprises.   There is  no doubt
that we  are in the  midst of a  revolutionary period of  discovery in
cosmology.

\section{WMAP Postscript}

Three months after we submitted this Colloquium article, the WMAP
Collaboration presented results from their first year of
data.\footnote{When the results were announced, cosmologists were
pleased to learn that the MAP satellite had been
re-named the Wilkinson Microwave Anisotrophy Probe (WMAP) to honor
David Wilkinson, a pioneer in the study of the CMB and a leader of the
MAP project, who died in September 2002.}  The results were at the
same time stunning and unsurprising. As several cosmologists put it, the
biggest surprise was the lack of a surprise.  Overlapping and
precision measurements have elevated cosmology to a new maturity,
where consistency is becoming a hallmark.

The angular power spectrum (cf Figure 9) was derived from five
all-sky maps with maximum angular resolution of
$0.2^\circ$ (30 times that of COBE) at frequencies
from 20 GHz to 100 GHz (reference).  The measurements were calibrated from the
Doppler shift of the CMB arising from Earth's motion around the
sun, $\delta T =(v/c)T_0 \simeq 0.27\,$mK ($v/c = 10^{-4}$).
WMAP's location a million miles from Earth helped keep systematics to
below 0.5\%.  From $\ell =2$ to $\ell \sim 350$ the measurements of
the multipole amplitudes were limited by sample (or cosmic) variance.
(Theories like inflation do not predict values for the individual
multipoles, but rather the variance of the distribution from which
they are draw.  The fact that for a given $\ell$ only $2\ell +1$
multipoles can be measured limits the precision with which the
variance can be estimated.)

The WMAP results (\onlinecite{Bennettetal:2003})
have sharpened and put on firmer footing
a large number of cosmological parameters (see Table I).  The
consistency of WMAP-determined parameters with previous
values was a strong indication of the increasing
reliability of cosmological results and their error estimates.
In particular, WMAP strengthened the case for dark matter
by its measurement of the ratio of the total amount of
matter to that in baryons, $\Omega_M/\Omega_B
= 6\pm 0.04$, and the case for dark energy
by showing that something like a cosmological
constant is needed to ``balance the books,''
$\Omega_X = 0.7 \pm 0.04$.  WMAP made clear that
our current consensus cosmology rests on a strong and
diverse, interlocking set of measurements.

While WMAP has yet to map CMB polarization (though it is
in the works), by detecting
the cross correlation between polarization and temperature
anisotropy it found the signature of the re-ionization
from UV starlight of the first stars at a redshift $z\simeq
20\pm 10$.  This is consistent with the predictions of
the CDM paradigm, and together with the SDSS quasars with redshifts
greater than 6 this now nicely brackets the re-ionization
history of the Universe:  at $z\sim 20$ the fraction of
free elections rose to around 50\% and by $z\simeq 6$ it
exceeded 99.99\%.

Two months after the WMAP results, a new compilation and analysis
of over 200 type Ia supernovae was presented (\onlinecite{Tonryetal:2003}),
and the direct evidence for cosmic
acceleration also grew stronger.  In particular, if dark energy is assumed
to have $w=-1$ (like a cosmological constant), the supernovae data
imply
\begin{eqnarray}
\Omega_X - 1.4\Omega_M & = & 0.35\pm 0.14 \nonumber \\
q_0 & = & -0.66 \pm 0.10 \nonumber \\
\Omega_X & = & 0.77 \pm 0.06 \qquad
{\rm assuming\ }\Omega_M =0.3 \pm 0.04   .
\end{eqnarray}

\bibliographystyle{apsrmp}


\vfill\eject


\begin{figure}
\epsfysize=5in
{\epsfbox{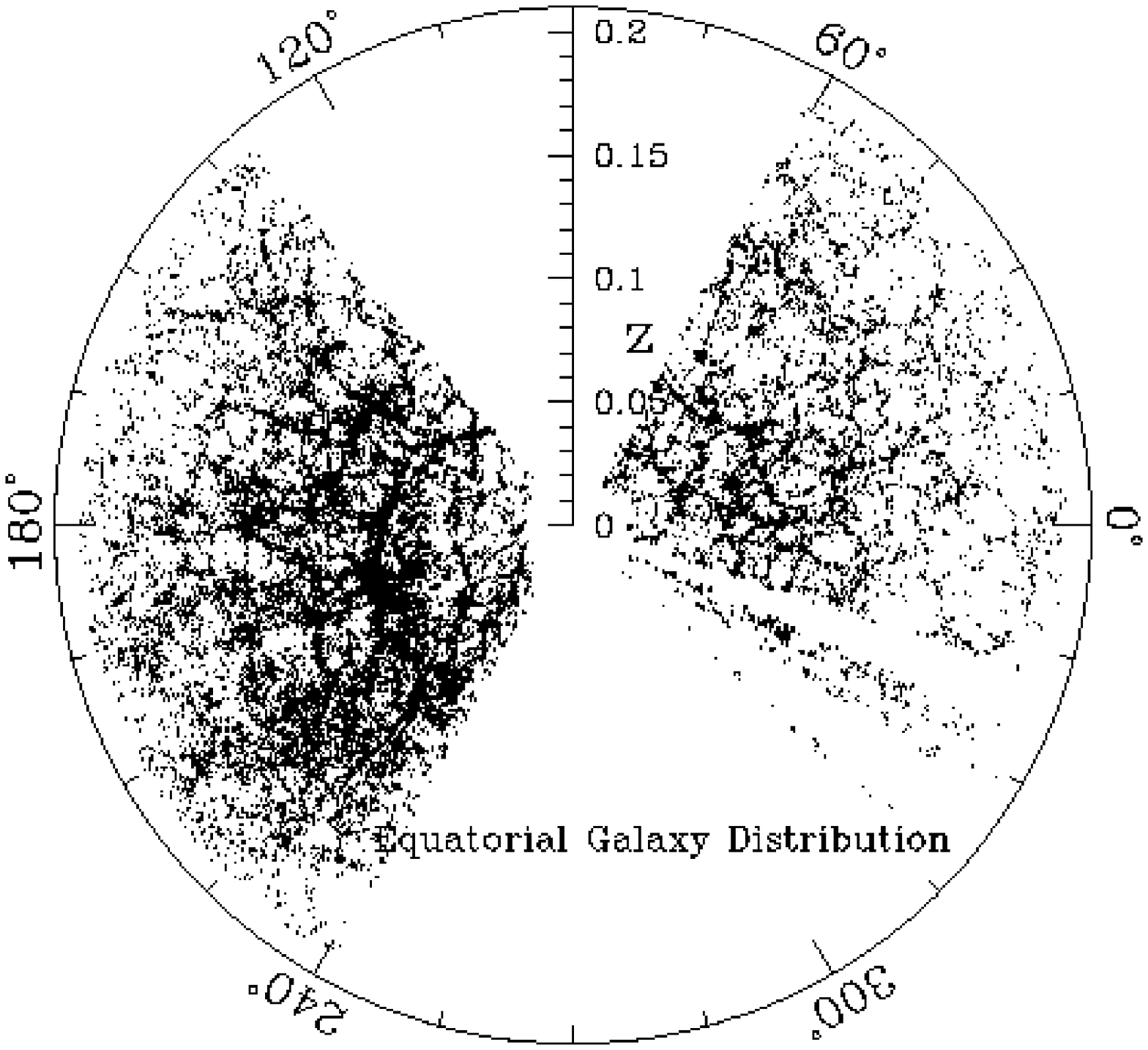} }
\caption{The Universe observed:
  A slice of the Universe constructed from the positions of
60,000 galaxies in the Sloan Digital Sky Survey.  Voids and walls can
be clearly seen (image courtesy of SDSS).
\label{1a}}
\end{figure}

\begin{figure}
\epsfysize=5in
{\epsfbox{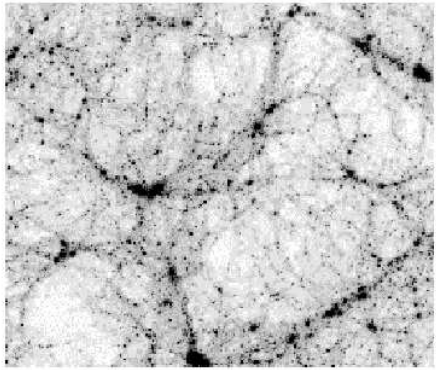}  }
\caption{The Universe  simulated:
 The distribution of dark matter in a large-scale
numerical simulation of the Universe.  The cosmic web of dark matter --
with its sheets, sinuous filaments and voids is apparent (image courtesy of
the Virgo Consortium). \label{1b}}
\end{figure}

\vfill\eject


\begin{figure}
\epsfysize=5in
\epsfbox{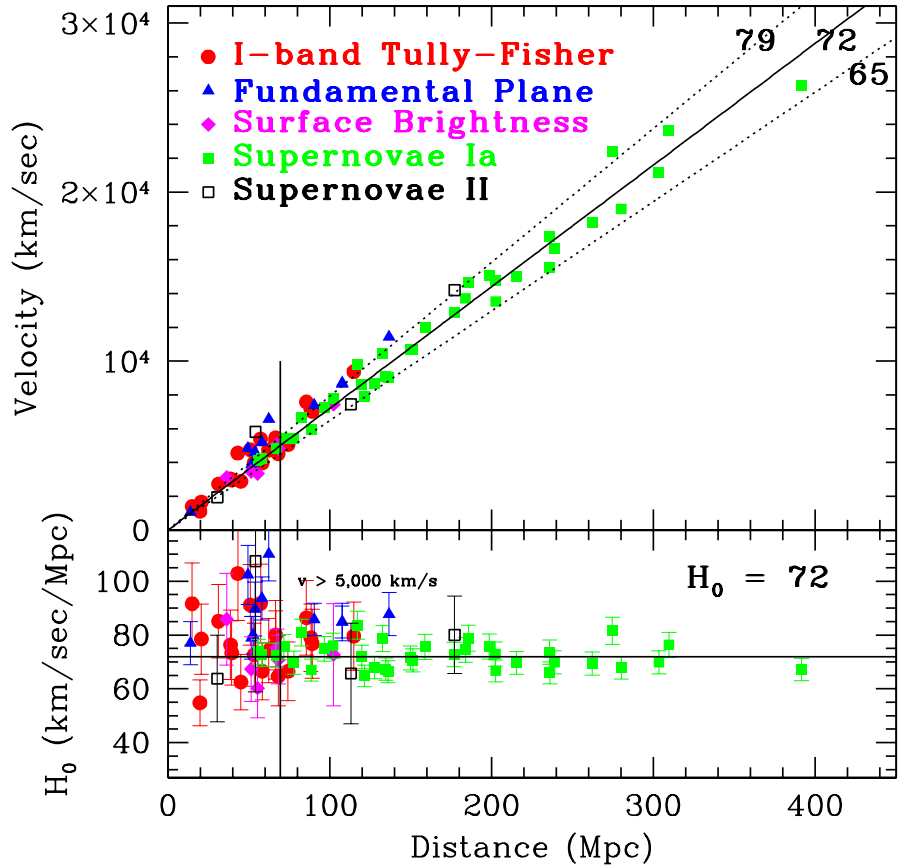}
\caption{Hubble diagram:   Low-redshift galaxies are used to
establish the expansion of the Universe and the Hubble constant; the
consistency of the five different distance indicators is shown.  The
lower panel shows the value of the Hubble constant object by object
and the convergence to 72\,km/s/Mpc. The scatter at distances less
than 100\,Mpc arises due to gravitational induced ``peculiar velocities''
that arise from the inhomogeneous distribution of matter.
\label{2a}}
\end{figure}

\vfill\eject

\begin{figure}
\epsfysize=5in
\epsfbox{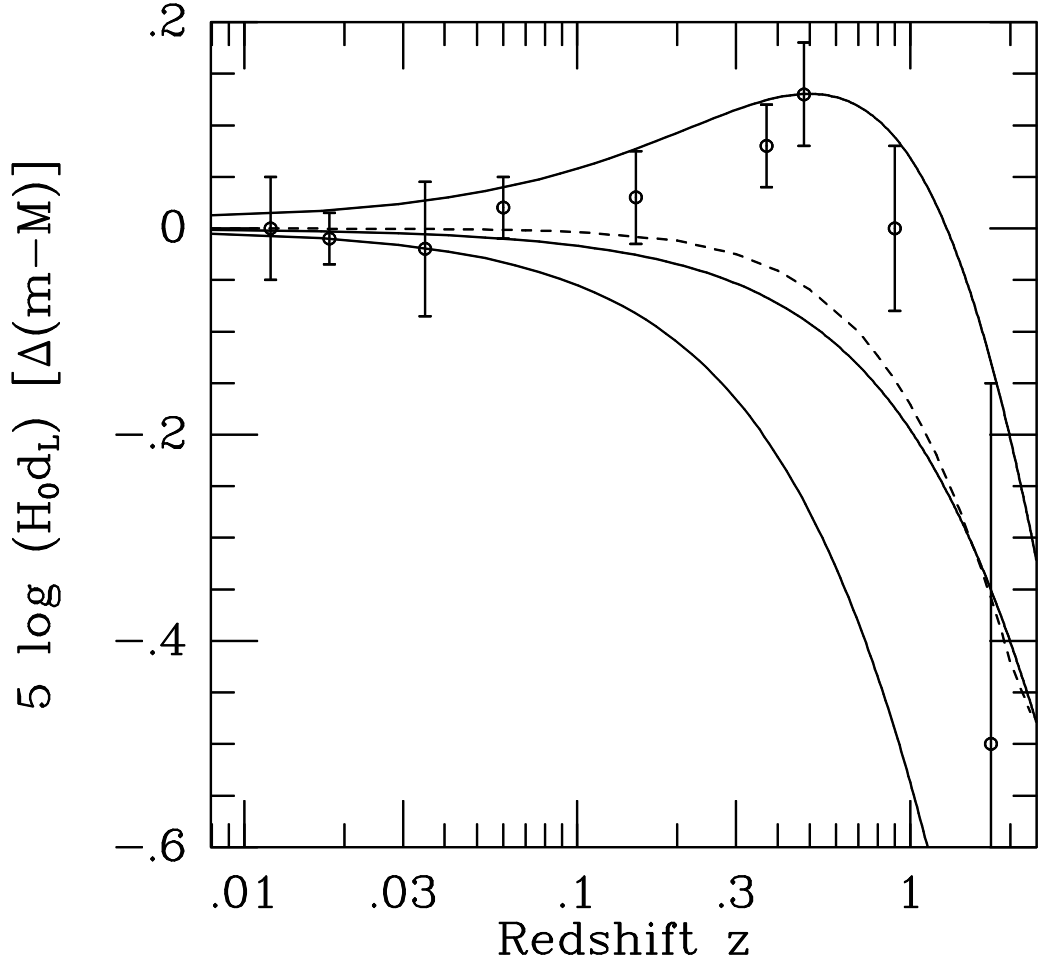}
\caption{Hubble diagram:  
 High-redshift type Ia supernovae probe the expansion
history and reveal accelerated expansion. In this differential
Hubble diagram the distance modulus, which is 5 times the logarithm
of the distance, relative to an empty Universe ($\Omega_0=0$) is
plotted.  Measurements from more than 200 type Ia supernova
are binned into 9 data points.  The solid curves represent three
theoretical models:
from the top, $\Omega_\Lambda = 0.7$ and $\Omega_M = 0.3$;
$\Omega_\Lambda = 0$ and $\Omega_M = 0.3$; and $\Omega_\Lambda
= 0$ and $\Omega_M = 1$.  The broken curve represents a nonaccelerating,
flat Universe (i.e., $q=0$ for all $z$); points above this curve
indicate acceleration (adapted from data in \onlinecite{Tonryetal:2003}).
\label{2b}}
\end{figure}

\vfill\eject

\begin{figure}
\epsfysize=5in
\epsfbox{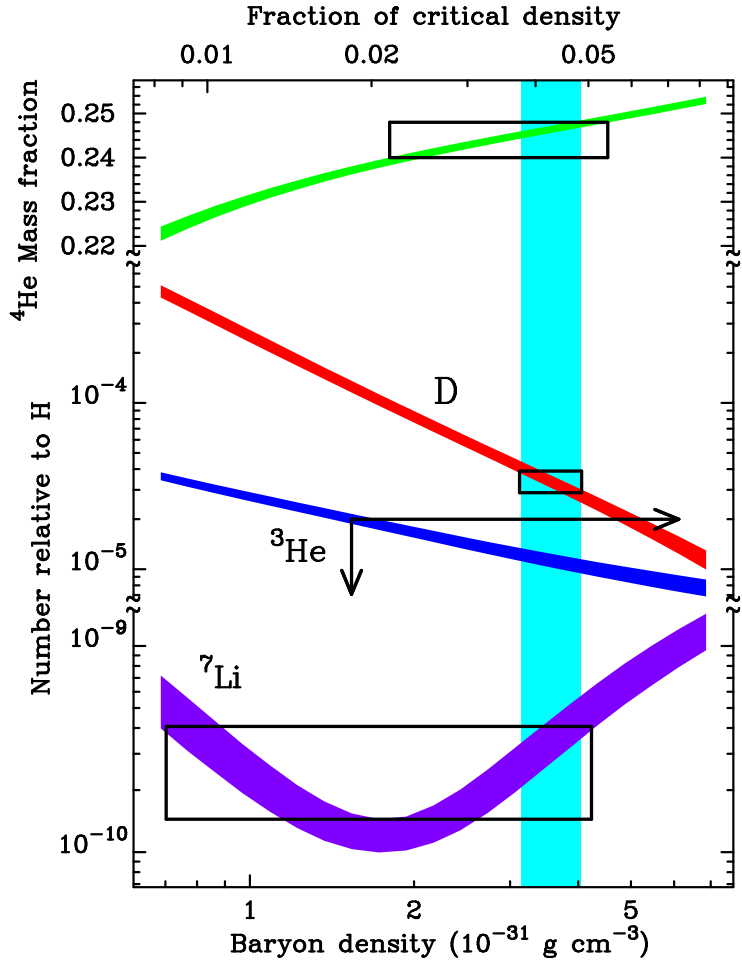}
\caption{The predicted abundance of the light elements vs. baryon
density.  The vertical band indicates the narrow range of baryon
densities consistent with the deuterium measurements; the boxes
(and open box for $^3$He) indicate the range in baryon density
(horizontal extent of box) that is consistent with the measured
light-element abundance (vertical extent of box).  The overlap
of the boxes with the deuterium band indicates the general consistency
of the observed abundances of the
other light elements with their predicted abundances for this baryon
density. (Note, for the $\Omega_B$ scale at the top, $h^2 = 0.5$
is assumed.) \label{3}}
\end{figure}

\vfill\eject

\begin{figure}
\epsfysize=5in
\epsfbox{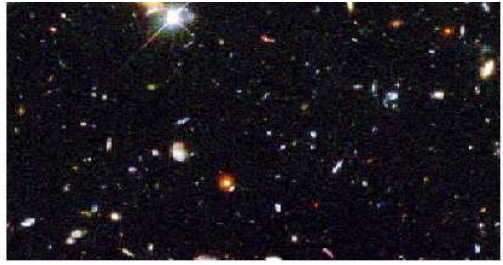}
\caption{The deepest image of the sky in visible light obtained
by the Hubble Space Telescope in 1995 (the Hubble Deep Field).  This
image revealed the time when typical galaxies (like our own Milky
Way) were forming (redshifts $z\sim 1-3$). In this image of about
one-forty-millionth of the sky, there is one star and more than
1500 galaxies (image courtesy of NASA). \label{4}}
\end{figure}

\vfill\eject


\begin{figure}
\epsfysize=5in
\epsfbox{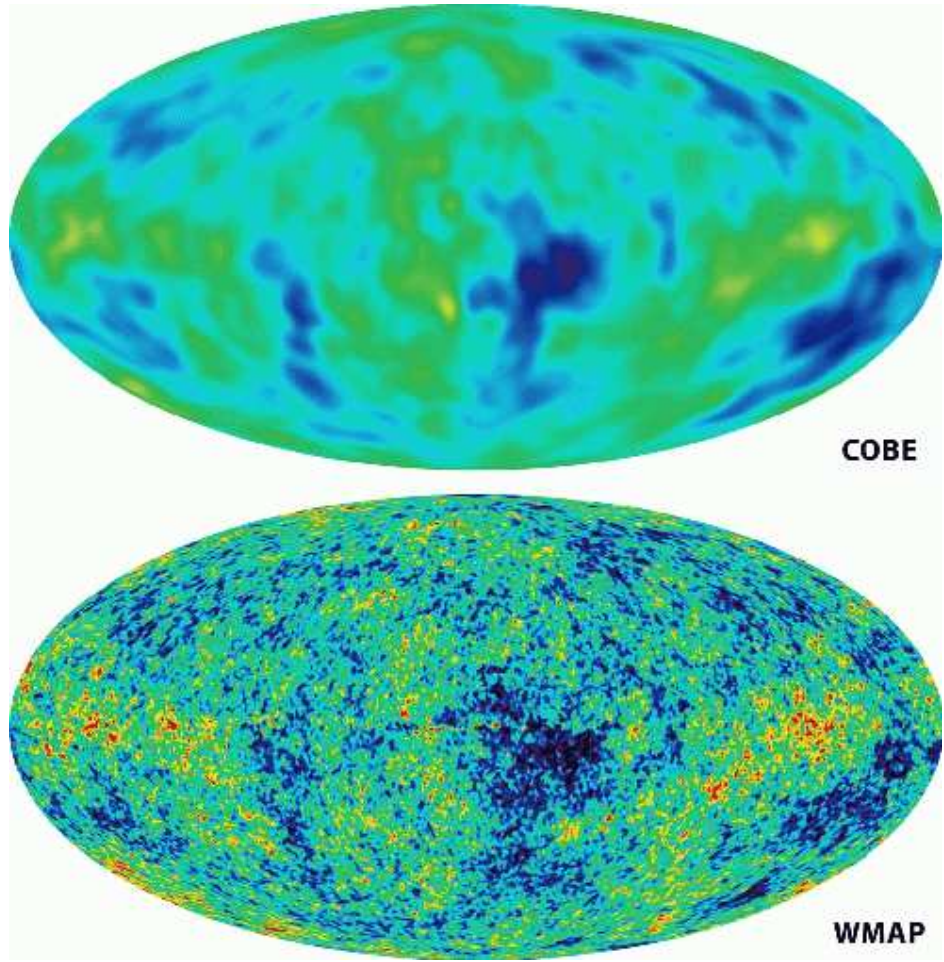}
\caption{Anisotropy of the Cosmic Microwave Background:
 All-sky maps, made by COBE (upper) and by WMAP
(lower); range of color scale is $\pm 200\mu$Kelvin.
The consistency of the 30 times higher resolution and
higher sensitivity WMAP results with COBE is apparent
(courtesy of NASA/WMAP Science Team). \label{5a}}
\end{figure}

\vfill\eject

\begin{figure}
\epsfysize=5in
\epsfbox{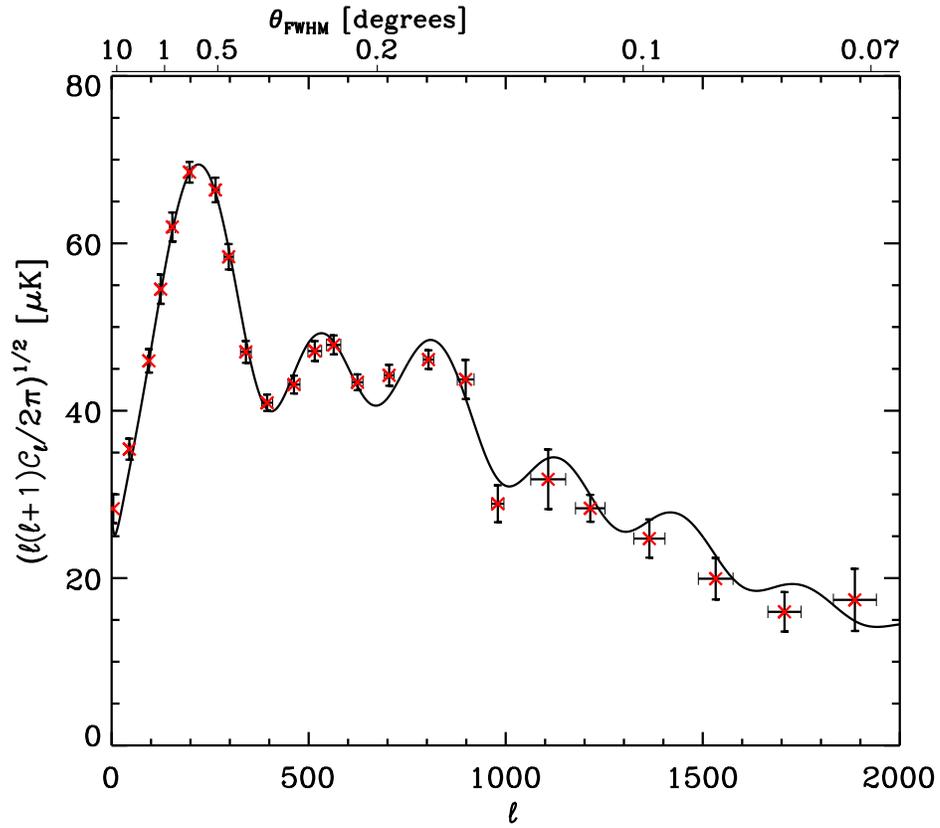}
\caption{Anisotropy of the Cosmic Microwave Background:
 Angular power spectrum of the CMB, incorporating
all the pre-WMAP data (COBE, BOOMERanG, MAXIMA, DASI, CBI,
ACBAR, FIRS, VSA, and other experiments).  Variance of the multipole
amplitude is plotted against multiple number; as indicated
by the top scale, multipole $\ell$ measures the fluctuations
on angular scale $\theta \sim 200^\circ /\ell$. Evidence of the
baryon -- photon oscillations can be seen as the
distinct ``acoustic peaks.'' The theoretical curve is the
consensus cosmological model (image courtesy of
C. Lineweaver). \label{5b}}
\end{figure}

\vfill\eject

\begin{figure}
\epsfysize=5in
\epsfbox{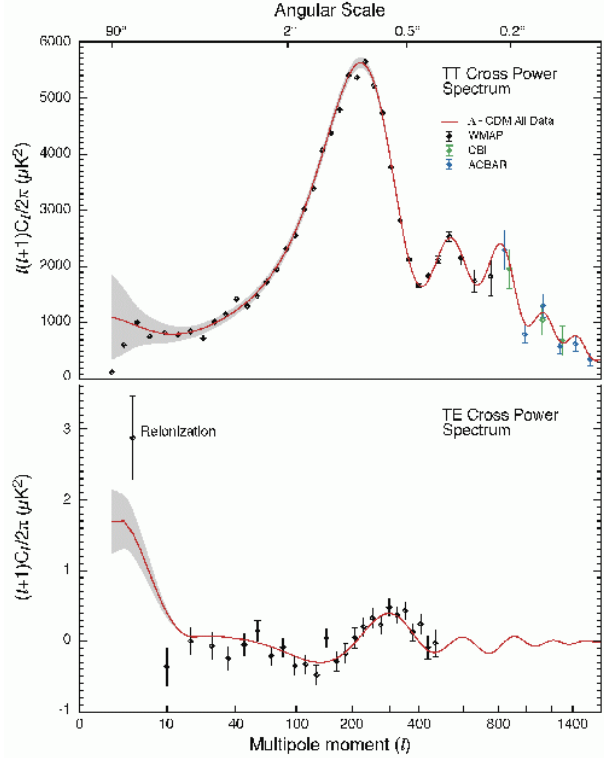}
\caption{Anisotropy of the Cosmic Microwave Background:
 The WMAP
angular power spectrum (also includes data from
CBI and ACBAR).  The curve is the consensus cosmology
model; the grey band includes cosmic variance.  The
WMAP measurements up to $\ell \sim 350$ are cosmic
variance limited. The lower panel shows the anisotropy
cross polarization power spectrum; the high point
marked re-ionization is the evidence for
re-ionization of the Universe at $z\sim 20$
 (courtesy of NASA/WMAP Science Team). \label{5c}}
\end{figure}

\vfill\eject

\begin{figure}
\epsfysize=5in
\epsfbox{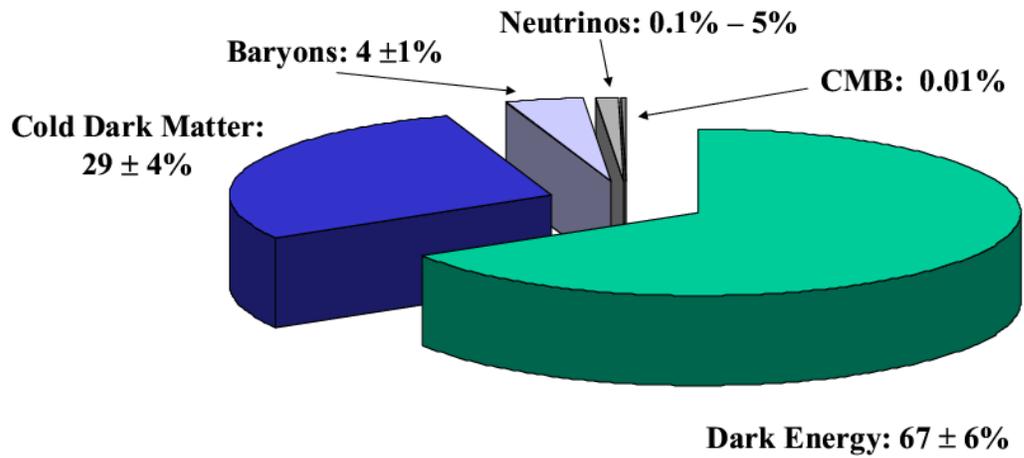}
\caption{The composition  of the Universe  today. Because the
different components of the mass/energy budget evolve differently,
the composition changes with time.  For example, at very
early times, photons and other relativitic particles were the
dominant component; from 10,000 years until a few billion years
ago, matter was the dominant component, and in the future dark
energy will be the dominant component. \label{6}}
\end{figure}

\vfill\eject

\begin{figure}
\epsfysize=5in
\epsfbox{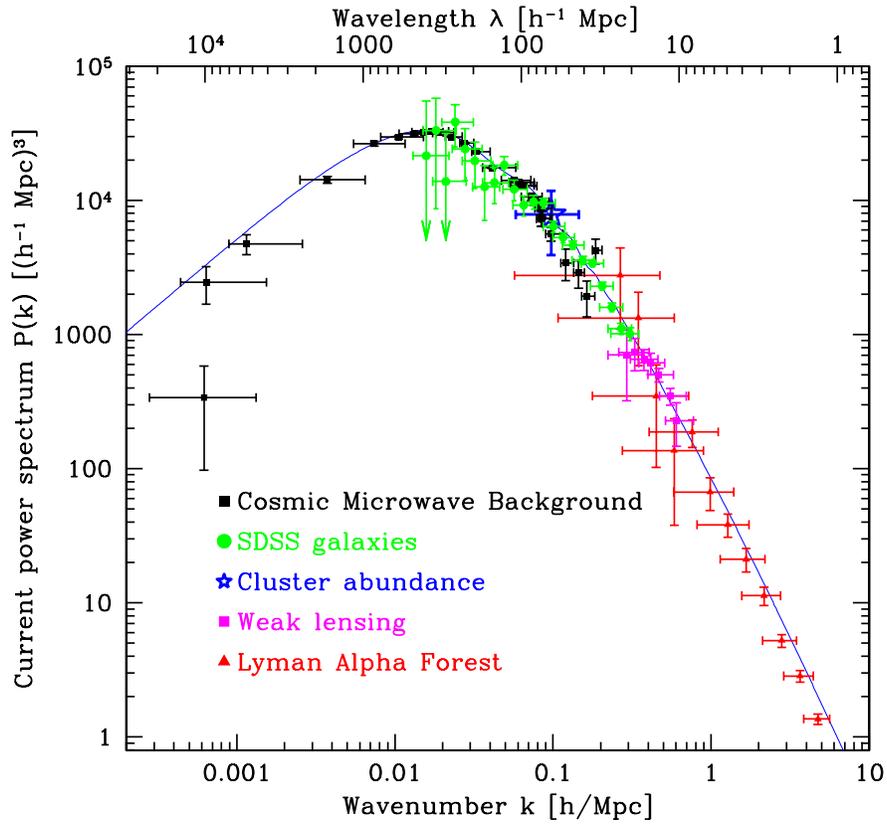}
\caption{Power spectrum of density inhomogeneity today obtained
from a variety of measurements including large-scale structure,
CMB, weak lensing,
rich clusters and the Lyman-alpha forest.  The curve is the
theoretical prediction for the consensus cosmology model
(from \onlinecite{Tegmark:2002}). \label{7}}
\end{figure}

\vfill\eject

\begin{figure}
\vspace{5in}
\end{figure}

\vfill\eject

\begin{table}
\caption{OUR 16 COSMOLOGICAL PARAMETERS}
\begin{tabular}{llll}
&&\\
\hline
\hline
Parameter & Value\footnotemark[1] & Description
& WMAP\footnotemark[2] \\
\hline
& & {\bf Ten Global Parameters} & \\
$h$ & $0.72 \pm 0.07$ & Present expansion rate\footnotemark[3]
& $0.71^{+0.04}_{-0.03}$ \\
$q_0$ & $-0.67\pm 0.25$ & Deceleration parameter\footnotemark[4]
& $-0.66\pm 0.10$\footnotemark[5] \\
$t_0$ & $13 \pm 1.5\,$Gyr & Age of the Universe\footnotemark[6]
& $13.7 \pm 0.2 \,$Gyr \\
$T_0$ & $2.725 \pm 0.001\,$K & CMB temperature\footnotemark[7]  \\
$\Omega_{0}$ & $1.03 \pm 0.03$ & Density parameter\footnotemark[8]
& $1.02\pm 0.02$ \\
$\Omega_{\rm B}$ & $0.039 \pm 0.008$ & Baryon Density\footnotemark[9]
& $0.044\pm 0.004$ \\
$\Omega_{\rm CDM}$ &$0.29 \pm 0.04$  & Cold Dark Matter Density\footnotemark[9]
& $0.23\pm 0.04$ \\
$\Omega_{\nu}$ & $0.001 - 0.05$  &  Massive Neutrino Density\footnotemark[10] \\
$\Omega_{X}$ & $0.67\pm 0.06$  & Dark Energy Density\footnotemark[9]
& $0.73 \pm 0.04$ \\
$w$ & $-1 \pm 0.2$  & Dark Energy Equation of State\footnotemark[11]
& $< -0.8\,$(95\% cl) \\
&&& \\
&& {\bf Six Fluctuation Parameters} & \\
$\sqrt{S}$   & $ 5.6^{+1.5}_{-1.0} \times 10^{-6} $
& Density Perturbation Amplitude\footnotemark[12] \\
$\sqrt{T}$    & $< \sqrt{S} $ & Gravity Wave Amplitude\footnotemark[13]
& $T < 0.71S\,$(95\%cl) \\
$\sigma_8$    & $0.9 \pm 0.1$  & Mass fluctuations on 8 Mpc\footnotemark[14]
& $0.84\pm 0.04$ \\
$n$   & $1.05\pm 0.09$ & Scalar index\footnotemark[8]
& $0.93\pm 0.03$\\
$n_T$ & --- &    Tensor index \\
$ dn/d\ln k$ & $-0.02\pm 0.04$ & Running of scalar index\footnotemark[15]
& $-0.03 \pm 0.02$ \\
&&& \\
\hline
&&& \\
\end{tabular}
\footnotetext[1]{\small The 1-$\sigma$ uncertainties quoted in
this table represent our combined analysis of published data.}
\footnotetext[2]{\small \onlinecite{Bennettetal:2003}.}
\footnotetext[3]{\small \onlinecite{Freedman:2001}; note:
$H_0 =100h\,{\rm km\,sec^{-1}\,Mpc^{-1}}$.}
\footnotetext[4]{\small Supernovae results combined with measurements
of the total matter density, $\Omega_M =\Omega_\nu + \Omega_B + \Omega_{\rm CDM}$
and $\Omega_0$, assuming $w=-1$ (\onlinecite{Perlmutter:1999,Riess:1998}).}
\footnotetext[5]{\small WMAP results (\onlinecite{Bennettetal:2003}) combined with
\onlinecite{Tonryetal:2003}.}
\footnotetext[6]{\small Value based upon CMB, globular cluster ages
and current expansion rate (\onlinecite{Knox:2001,KraussChaboyer:2002,Oswald:1996}).}
\footnotetext[7]{\small \onlinecite{Mather:1999}.}
\footnotetext[8]{\small Combined analysis of four CMB measurements
(\onlinecite{Sievers:2002}).}
\footnotetext[9]{\small Combined analysis of CMB, BBN, $H_0$ and
cluster baryon fraction (\onlinecite{Turner:2002}).}
\footnotetext[10]{\small Lower limit from SuperKamiokande measurements;
upper limit from structure formation (\onlinecite{Fukuda:2002,Elgaroy:2002}).}
\footnotetext[11]{\small Supernova measurements, CMB and large-scale
structure (\onlinecite{PerlmutterTurnerWhite:1999}).}
\footnotetext[12]{\small Contribution of density perturbations to the
variance of the CMB quadrupole (with $T$=0) (\onlinecite{Gorski:1996}).}
\footnotetext[13]{\small Contribution of gravity waves to the
variance of the CMB quadrupole (upper limit) (\onlinecite{Kinney:2001}).}
\footnotetext[14]{\small Variance in values reported is larger
than the estimated errors; adopted error reflects this (\onlinecite{Lahav:2002}).}
\footnotetext[15]{\small Deviation of the scalar perturbations from a
pure power law (\onlinecite{LewisBridle:2002}).}
\end{table}

\end{document}